\documentclass[11pt]{article}
\usepackage{amsmath}
\usepackage{pstricks}

\def\half{\frac{1}{2}}
\def\bsh{\backslash}

\newfont{\bbbold}{msbm10 scaled \magstep1}

\def\bbF{\mbox{\bbbold F}}

\def\bbR{\mbox{\bbbold R}}

\def\cD{{\cal D}}
\def\cE{{\cal E}}
\def\cF{{\cal F}}

\def\cL{{\cal L}}

\def\cN{{\cal N}}
\def\cO{{\cal O}}

\def\cR{{\cal R}}

\def\cV{{\cal V}}

\def\cX{{\cal X}}

\def\cX{{\cal X}}

\newfont{\goth}{eufm10 scaled \magstep1}

\def\ge{\mbox{\goth e}}

\def\gg{\mbox{\goth g}}

\def\gi{\mbox{\goth i}}

\def\gn{\mbox{\goth n}}
\def\go{\mbox{\goth o}}
\def\gp{\mbox{\goth p}}

\def\gs{\mbox{\goth s}}

\def\a{\alpha}\def\adt{\dot \alpha}
\def\b{\beta}
\def\c{\gamma}\def\C{\Gamma}\def\cdt{\dot\gamma}
\def\d{\delta}
\def\ve{\varepsilon}
\def\f{\phi}\def\F{\Phi}
\def\h{\eta}

\def\L{\Lambda}

\def\S{\Sigma}

\def\th{\theta}\def\Th{\Theta}

\def\be{\begin{equation}}\def\ee{\end{equation}}
\def\bea{\begin{eqnarray}}\def\eea{\end{eqnarray}}
\def\barr{\begin{array}}\def\earr{\end{array}}

\def\o{\omega}\def\O{\Omega}
\def\del{\partial}
\def\ua{\underline{\alpha}}
\def\ub{\underline{\phantom{\alpha}}\!\!\!\beta}
\def\uc{\underline{\phantom{\alpha}}\!\!\!\gamma}

\def\um{\underline{\mu}}
\def\ud{\underline\delta}
\def\ue{\underline\epsilon}

\def\wt{\widetilde}

\def\xz{\times}

\let\la=\label

\def\nn{\nonumber}
\def\bd{\begin{document}}
\def\ed{\end{document}}
\def\ba{\begin{array}}
\def\ea{\end{array}}
\def\bea{\begin{eqnarray}}
\def\eea{\end{eqnarray}}
\def\ft#1#2{\tfrac{#1}{#2}}
\def\fft#1#2{\frac{#1}{#2}}
\def\sst#1{{\scriptscriptstyle #1}}
\def\oneone{\rlap 1\mkern4mu{\rm l}}

\newcommand{\eq}[1]{(\ref{#1})}
\newcommand{\w}[1]{\\[0.#1cm]}
\def\eqs#1#2{(\ref{#1}-\ref{#2})}
\def\det{{\rm det\,}}
\def\tr{{\rm tr}}

\newcommand{\hoch}[1]{$\, ^{#1}$}
\newcommand{\imperial}{\it\small Theoretical Physics Group, Imperial College London\\ Prince Consort Road, London SW7 2AZ, UK}
\newcommand{\kings}
{\it\small Department of Mathematics, King's College, University of London\\ Strand, London WC2R 2LS, UK}
\newcommand{\uu}
{\it\small Department of Theoretical Physics, Uppsala, Sweden}
\newcommand{\hip}
{\it\small HIP-Helsinki Institute of Physics, P.O. Box 64 FIN-00014
University of Helsinki, Suomi-Finland}
\newcommand{\stock}
{\it\small Department of Theoretical Physics, Stockholm, Sweden}
\newcommand{\golm}
{\it\small AEI, Max Planck Institut f\"ur Gravitationsphysik\\ Am M\"{u}hlenberg 1, D-14476 Potsdam, Germany}
\makeatletter
\renewcommand\theequation{\thesection.\arabic{equation}}
\@addtoreset{equation}{section} \makeatother

\newcommand{\sa}{/ \hspace{-1.2ex}}
\newcommand{\saa}{/ \hspace{-1.4ex}}
\newcommand{\saaa}{\, / \hspace{-1.6ex}}
\newcommand{\Scal}[1]{\Bigl ({#1} \Bigr )}
\newcommand{\scal}[1]{\bigl ({#1} \bigr )}

\newcommand{\CR}{\nonumber \\*}

\newcommand{\trace}{\hbox {tr}~}
\newcommand{\traceS}{\hbox {tr}_{\scriptscriptstyle \mathfrak{S}}~}

\DeclareMathAlphabet{\mathpzc}{OT1}{pzc}{m}{it}
\def\BRST{\,\mathpzc{s}\,}
\def\aBRST{{\scriptstyle (\mathpzc{s})}}
\def\q{{{\scriptscriptstyle (Q)}}}
\def\qs{{\scriptscriptstyle (Q\mathpzc{s})}}
\def\Qsla{{\mathcal{S}_{\q}}}
\def\Slav{{\mathcal{S}_\aBRST}}
\def\epsilonb{{\overline{\epsilon}}}
\def\bulletup{{\scriptstyle \bullet}}

\newcommand{\gra}[2]{{\scriptscriptstyle (#1 , #2 )}}
\newcommand{\ord}[1]{{\scriptscriptstyle (#1)}}

\def\cL{{\cal L}}
\def\cN{\mathcal{N}}
\def\cO{\mathcal{O}}

\def\ie{{\it i.e.}\ }
\def\eg{{\it e.g.}\ }

\newcommand{\sfrac}[2]{{\scriptstyle \frac{#1}{#2}}}
\newcommand{\stfrac}[2]{{\scriptscriptstyle \frac{#1}{#2}}}

 \def\balpha{{\overline{\alpha}}}
 \def\bbeta{{\overline{\beta}}}
 \def\bgamma{{\overline{\gamma}}}
 \def\bdelta{{\overline{\delta}}}
 \def\bepsilon{{\overline{\epsilon}}}
 \def\bvarepsilon{{\overline{\varepsilon}}}
 \def\bzeta{{\overline{\zeta}}}
 \def\bareta{{\overline{\eta}}}
 \def\btheta{{\overline{\theta}}}
 \def\bvartheta{{\overline{\vartheta}}}
 \def\biota{{\overline{\iota}}}
 \def\bkappa{{\overline{\kappa}}}
 \def\blambda{{\overline{\lambda}}}
 \def\bmu{{\overline{\mu}}}
 \def\bnu{{\overline{\nu}}}
 \def\bxi{{\overline{\xi}}}
 \def\bpi{{\overline{\pi}}}
 \def\brho{{\overline{\rho}}}
 \def\bvarrho{{\overline{\varrho}}}
 \def\bsigma{{\overline{\sigma}}}
 \def\bvarsigma{{\overline{\varsigma}}}
 \def\btau{{\overline{\tau}}}
 \def\bphi{{\overline{\phi}}}
 \def\bvarphi{{\overline{\varphi}}}
 \def\bchi{{\overline{\chi}}}
 \def\bpsi{{\overline{\psi}}}
 \def\bomega{{\overline{\omega}}}

\def\thalf{{\textrm{\tiny\textonehalf}}}
\def\tquarter{{\textrm{\tiny\textonequarter}}}
\def\Ko{{\scriptscriptstyle K}}
\def\tKo{\scriptscriptstyle k }
\def\corr{$\clubsuit$}

\newcommand{\auth}{\large J. Greitz\footnote{email:jesper.greitz@kcl.ac.uk}, P.S.\ Howe\footnote{email: paul.howe@kcl.ac.uk}}

\begin{document}

\renewcommand{\thefootnote}{\fnsymbol{footnote}}

\null
\begin{flushright}
{\small KCL-MTH-11-07}\\
\vskip 1.5 cm
\end{flushright}

\begin{center}
{\Large{\bf Maximal supergravity in three dimensions: supergeometry and differential forms}}
\vspace{.75cm}

\auth

\vspace{.5cm}

\begin{center}
{\it\small Department of Mathematics, King's College, London, UK}
\end{center}
\vspace{1cm}

{\bf Abstract}
\end{center}
\vskip .5cm
The maximal supergravity theory in three dimensions, which has local $SO(16)$ and rigid $E_8$ symmetries, is discussed in a superspace setting starting from an off-shell superconformal structure. The on-shell theory is obtained by imposing further constraints. It is essentially a non-linear sigma model that induces a Poincar\'e supergeometry that is described in detail. 
The possible $p$-form field strengths, for $p=2,3,4$, are explicitly constructed using supersymmetry and $E_8$. The gauged theory is also discussed.

\vspace{1cm}


\renewcommand{\thefootnote}{\arabic{footnote}}
\setcounter{footnote}{0}

\pagebreak
\tableofcontents
\setcounter{page}{1}


\section{Introduction}


It has been known for many years that maximal supergravity theories have hidden rigid symmetry groups that  increase in dimension as the dimension of spacetime decreases \cite{Cremmer:1979up}. The $D=3$ case is special in the sense that the symmetry group $E_8$ is the largest finite one in the $E$ series; in $D=2$ one has $E_9$ \cite{Nicolai:1988jb}. More recently it has been suggested that these symmetries might be extended to $E_{10}$ \cite{Kleinschmidt:2004dy} or $E_{11}$ \cite{West:2001as}. In addition, in the $D=3$ theory \cite{Marcus:1983hb}, the $128+128$ on-shell degrees of freedom are entirely non-gravitational, so that in a sense the theory is really an $SO(16)\bsh E_8$ non-linear sigma model, although there is an induced geometrical structure.

In this paper we describe this model in a superspace with a local $SL(2,\bbR)\xz SO(16)$ structure group. We begin with a brief review of the geometry that describes the off-shell superconformal multiplet and then impose further constraints to accommodate the sigma model fields. Because the scalars transform non-linearly under $E_8$ and the fermions transform according to a spinor representation of $SO(16)$ it follows that the former can only appear covered by derivatives in the geometrical tensors, and that the latter cannot appear linearly. This circumstance simplifies the geometry somewhat, especially due to the fact that the dimension one-half torsion tensor must vanish which is not the case in higher dimensions (except $D=11$). This might lead one to believe that there could be generalised chiral (CR) structures involving more mutually anti-commuting odd derivatives than in higher-dimensional spacetimes, but this is not the case due to obstructions arising from the curvature.

In recent years there have been several studies of the systematics of form fields in supergravity theories, starting with \cite{Cremmer:1997ct,Cremmer:1998px}. It was realised that this could be formalised in terms of Borcherds algebras \cite{HenryLabordere:2002dk}, and also in terms of $E_{11}$ \cite{Julia:1997cy,Riccioni:2007au,Bergshoeff:2007qi,Riccioni:2009xr}. (See \cite{Henneaux:2010ys} for a discussion of the relation between the two). In a separate, but related development, it has been shown that the same sets of forms contribute to the hierarchies found in gauged supergravities, see \cite{deWit:2008gc} and references therein. A key feature is that these forms fall into representations of the duality groups. In addition to the physical forms and their duals there are also $(D-1)$-form potentials, related to gaugings, and $D$-form potentials, related to space-filling branes. We discuss these fields explicitly and show that all of the coupled Bianchi identities for the associated field strengths are satisfied.  Our construction uses only supersymmetry and $E_8$ symmetry, although the allowed representations that the forms transform under agree with those predicted from $E_{11}$. The superspace method has some advantages, especially for the top forms (three-form potentials in $D=3$). This is because it makes sense to consider four- (and indeed higher)-form field strengths in superspace due to the fact that the odd basis differential forms are commutative. This point of view was advocated previously in the context of maximal supersymmetry in ten dimensions \cite{Bergshoeff:2007ma,Bergshoeff:2010mv}. An additional feature of the formalism is that it is manifestly supersymmetric, and indeed, if one concentrates on the field strengths, manifestly covariant under all symmetries.

Three dimensions is special for gauged maximal supergravity \cite{Nicolai:2000sc,Nicolai:2001sv,deWit:2008ta} because there are no independent vector degrees of freedom. Gauging is very natural in superspace and the constraints on the embedding tensor can be seen as a direct consequence of the gauged Maurer-Cartan equation. Essentially, the dimension-one constraints are modified by a set of functions that fit together in the $1+3875$ representations of $E_8$. The higher-rank forms, discussed in detail in \cite{deWit:2008ta}, can also be deformed, and we choose to do this in a covariant fashion, i.e. by deforming the Bianchi identities. Discussions of the $E_{11}$ approach to gauged supergravity in $D=3$ can be found in \cite{Bergshoeff:2007qi,Bergshoeff:2008qd}.

The organisation of the paper is as follows: in section 2 we describe the geometrical set-up and review the off-shell superconformal constraints for $N$-extended supergravity in $D=3$. In section 3 we introduce the $E_8$ sigma model in the context of this supergravity background and show how the latter can accommodate it by making appropriate identifications. In section 4 we analyse the geometrical Bianchi identities up to dimension two. This allows us to identify all components of the torsion and curvature tensors and to verify explicitly that the constraints are consistent. It is important to do this as the initial set of constraints include conventional ones for the $SO(16)$ connection, whereas the sigma model fixes this connection in terms of the physical fields, so that one needs to check that these choices are compatible.  In section 5 we turn our attention to the additional form fields. These consist of the duals to the scalars (two-form field strengths) as well as three- and four-forms whose potentials have no physical degrees of freedom, and we also give a brief discussion of possible five-forms. The analysis of the associated Bianchi identities is considerably simplified by the use of superspace cohomology. In section 6 we discuss the gauging of the theory in a superspace setting. We start by modifying the geometry by gauging the Maurer-Cartan equation and then deform the Bianchi identities for the form fields. In the final section we make some concluding remarks. There are appendices on our conventions, superspace cohomology, some details of a calculation, and a final one in which we show, using harmonic superspace, that there are no higher-rank (Lorentzian) CR structures in this theory than in higher dimensional maximal supergravity.

\section{Geometrical set-up}


We consider a supermanifold $M$ with (even$|$odd)-dimension $(3|32)$. The basic structure is determined by a choice of odd tangent bundle $T_1$ such that the Frobenius tensor, which maps pairs of sections of $T_1$ to the even tangent bundle, $T_0$, generates the latter. We shall also suppose that there is a preferred basis $E_{\a i},\,\a=1,2;\,i=1,\ldots 16$ for $T_1$ such that the components of the Frobenius tensor, which we shall also refer to as the dimension-zero torsion, are

\be
 T_{\a i\b j}{}^c=-i\d_{ij} (\c^c)_{\a\b}\,;\qquad c=0,1,2\ .
 \la{2.1}
\ee

At this stage $T_0$ is defined as the quotient, $T/T_1$, but we can make a definite choice for $T_0$ by imposing some suitable dimension one-half constraint. When this has been done, the structure group will be reduced to $SL(2,\bbR)\xz SO(16)$, with the Lorentz vector indices being acted on by the local  $SO(1,2)$ associated with $SL(2,\bbR)$.\footnote{The dimension-zero torsion \eq{2.1} is also invariant under local Weyl rescalings, but we shall not include this factor in the structure group.  It is broken in the on-shell Poincar\'e theory.} With respect to this structure we have preferred basis vector fields $E_A=(E_a,E_{\ua})=(E_a,E_{\a i})$ with dual one-forms $E^A=(E^a,E^{\ua})=(E^a, E^{\a i})$, the latter being related to the coordinate basis forms $dz^M=(dx^m, d\th^{\um})$ by the supervielbein matrix $E_M{}^A$, i.e. $E^A=dz^M E_M{}^A$. Here, coordinate indices are taken from the middle of the alphabet, preferred basis indices from the beginning, while even (odd) indices are latin and greek respectively. Underlined odd indices run from 1 to 32, and $SO(16)$ vector indices are denoted $i,j$ etc. 

We now introduce a set of connection one-forms, $\O_A{}^B$, for the above structure group. We have

\bea
 \O_a{}^{\ub}&=& \O_{\ua}{}^b=0\nn\w1
 \O_{\a i}{}^{\b j}&=& \d_i{}^j \O_\a{}^\b + \d_\a{}^\b \O_i{}^j\nn\w1
 \O_a{}^b&=& - (\c_a{}^b)_\a{}^\b \O_\b{}^\a\ .
 \la{2.2}
\eea

Spinor indices $\a,\b$ are raised and lowered by the epsilon tensor, while Lorentz and $SO(16)$ vector indices are raised by the corresponding metrics $\h_{ab}, \d_{ij}$. We have $\O_{\a\b}=\O_{\b\a}$ while $\O_{ab}$ and $\O_{ij}$ are antisymmetric. The torsion and curvature are defined in the usual way

\bea
 T^A&=& DE^A:=d E^A + E^B \O_B{}^A\nn\w1
 R_A{}^B&=& d\O_A{}^B + \O_A{}^C \O_C{}^B\ .
 \la{2.3}
\eea

The Bianchi identities are

\bea
 DT^A&=& E^B R_B{}A\nn\w1
 D R_A{}^B&=&0\ .
 \la{2.4}
\eea

Equation \eq{2.1} does not simply determine the structure group, it is also a constraint. With an appropriate choice of dimension one-half connections and of $T_0$, and making use of the dimension one-half Bianchi identity, one finds that all components of the dimension one-half torsion may be set to zero:

\be
 T_{\ua\ub}{}^{\uc}=T_{a \ub}{}^c=0\ .
 \la{2.5}
\ee

Imposing further conventional constraints corresponding to the dimension-one connection components we find that the dimension-one torsion can be chosen to have the form

\bea
 T_{ab}{}^c&=&0\nn\w1
 T_{a \b j}{}^{\c k}&=&   (\c_a)_\b{}^\c K_j{}^k + (\c^b)_\b{}^\c L_{ab j}{}^k\ ,
 \la{2.6}
\eea

where $K_{ij}$ is symmetric and $L_{abij}$ is antisymmetric on both pairs of indices. The dimension-one curvatures are

\bea
 R_{\a i\b j, cd}&=& -2i(\c_{cd})_{\a\b} K_{ij} -2i\ve_{\a\b} L_{cdij}\nn\w1
 R_{\a i\b j,kl}&=& i\ve_{\a\b}(M_{ijkl} + 4 \d_{[i[k} K_{j]l]})-i(\c^a)_{\a\b} (4\d_{(i[k} L_{a j) l]}-\d_{ij} L_{a kl})\ ,
 \la{2.7}
\eea

where $L_{ab}=\ve_{abc} L^c$, and $M_{ijkl}$ is totally antisymmetric. This geometry is valid for any $N$, not just $N=16$, and describes an off-shell superconformal multiplet \cite{Howe:1995zm}. The interpretation of the dimension-one fields, $K,L,M$, is as follows. The geometry is determined by the basic constraint \eq{2.1} which is invariant under Weyl rescalings where the parameter is an unconstrained scalar superfield. This means that some of the fields that appear in the geometry do not belong to the Weyl supergravity multiplet. At dimension one $K$ and $L$ are of this type, so that we could set them to zero if we were only interested in the superconformal multiplet. The field $M_{ijkl}$, on the other hand, can be considered as the field strength superfield for the Weyl supermultiplet \cite{Howe:1995zm}.\footnote{This was discussed explicitly in  for the case of $N=8$ in \cite{Howe:2004ib}.} The fact that $M$ is not expressible in terms of  the torsion is due to a lacuna in Dragon's theorem  \cite{Kuzenko:2011xg,Cederwall:2011pu} which in higher-dimensional spacetimes states that the curvature is so determined \cite{Dragon:1978nf}. We recall that in three-dimensional spacetime there is no Weyl tensor but that its place is taken by the dimension-three Cotton tensor. This turns out to be a component of the superfield $M_{ijkl}$ so that we could refer to the latter as the super Cotton tensor. Using the notation $[k,l]$ to denote fields that have $k$ antisymmetrised $SO(N)$ indices and $l$ symmetrised spinor indices, one can see that the component fields of the superconformal multiplet fall into two sequences starting from $M$. The first has fields of the type $[4-p,p]$, where the top ($[4,0]$) component is the supersymmetric Cotton tensor, while the second has fields of the type $(4+p,p)$ and therefore includes higher spin fields for $N>8$. There is also a second scalar $(4,0)$ at dimension two. Fields with two or more spinor indices obey covariant conservation conditions so that each field in the multiplet has two degrees of freedom multiplied by the dimension of the $SO(N)$ representation, provided that we count the dimension one and two scalars together. It is easy to see that the number of bosonic and fermionic degrees of freedom in this multiplet match.

The other components of the curvature and torsion can be derived straightforwardly from here, but we shall postpone this until we have introduced the physical fields. For the conformal case, the dimension three-halves Bianchi identities were solved explicitly in \cite{Kuzenko:2011xg}, while a  detailed discussion of the $N=8$ case has been given in \cite{Cederwall:2011pu}.


\section{The sigma model}


As we have seen the structure group contains an $SO(16)$ factor which is associated with an $SO(16)$ principal bundle. The sigma model can be introduced via  the requirement that this bundle can be lifted to a flat $E_8$ bundle. Our conventions for the Lie algebra, $\ge_8$, are as follows: the generators are $(M_{ij}, N_I)$ where $M_{ij}=-M_{ji}$ are the generators for $\gs\go(16)$ and the remaining generators, $N_I,\,I=1,\ldots 128$, transform under one of the two Weyl spinor representations of $\gs\gp\gi\gn(16)$. We shall denote the other representation by primed indices, e.g.  $I'$. The algebra of $\ge_8$ is

\bea
 [M_{ij}, M^{kl}]&=& -4\d_{[i}{}^{[k} M_{j]}{}^{l]}\nn\w1
 [M_{ij},N_I]&=& -\half(\S_{ij})_{IJ} N_J\nn\w1
 [N_I, N_J]&=& (\S^{ij})_{IJ} M_{ij}\ ,
 \la{3.1}
\eea

where the $SO(16)$ sigma matrices are denoted by $\S$ (see appendix for conventions).
The sigma model field can be viewed as a section $\cV$ of the $E_8$ bundle. It is acted on to the right by $E_8$ and to the left by the local $SO(16)$ and therefore corresponds to an $SO(16)\bsh E_8$ sigma model superfield. The Maurer-Cartan form is

\be
 \F:=d\cV \cV^{-1}:=P+ Q\ ,
 \la{3.2}
\ee

where $Q=\half \O^{ij} M_{ij}$ and where $P$ takes its values in the quotient algebra, i.e. $P=P^I N_I$. From the Maurer-Cartan equation (vanishing $E_8$ curvature), $d\F+\F^2=0$, we find

\bea
 DP&=&0\w1
 R&=&-P^2\ ,
 \la{3.3}
\eea

where $R:= \half R^{ij} M_{ij}$ is the $SO(16)$ curvature, while $D$ is the $SO(16)$-covariant exterior derivative. In indices, the above equations are

\bea
 2 D_{[A} P_{B]} + T_{AB}{}^C P_C&=& 0\w1
 \la{3.4}
 R_{AB,ij}&=&  2P_A \S_{ij} P_B\ .
 \la{3.5}
\eea

We shall need to impose a constraint on the dimension one-half component of $P$ to ensure that we have the correct number of degrees of freedom, namely 128 bosonic and fermionic. We therefore set

\be
 P_{\a i I}=i(\S_i \L_\a)_I\ ,
 \la{3.6}
\ee

where $\L_{\a I'}$ describes the physical 128 spin one-half fields. The dimension-one component of \eq{3.4} is then satisfied if

\be
 D_{\a i} \L_{\b I'}=\half(\c^a)_{\a\b} (\S_i P_a)_{I'}\ .
 \la{3.7}
\ee

We can think of $P_{a I}$ as essentially the spacetime derivative of the physical scalar fields. In order to see this more explicitly, it is perhaps useful to look at the linearised limit. In the physical gauge we can put $\cV=\exp (\f^I N_I)$ where $\f^I$ denotes the 128 scalars. If we now keep only terms linear in the fields we find

\bea
 D_{\a i} \f_I&=& i (\S_i \L_\a)_I \nn\w1
 D_{\a i}\L_{\b J'}&=& \half(\c^a)_{\a\b} (\S_i P_a)_{J'}=\half(\c^a)_{\a\b} (\S_i \del_a \f)_{J'}\ ,
 \la{3.7.1}
\eea

where $D_{\a i}$ is now the usual supercovariant derivative in flat superspace. It follows from \eq{3.7.1} that both $\f_I$ and $\L_{\a I'}$ satisfy free field equations of motion.

Note that we have now specified the $SO(16)$ connection in two ways, by choosing corresponding conventional constraints on the torsion, and explicitly in terms of $\cV$. We therefore need to verify that these are compatible. We can easily see that they are by making use of the dimension-one component of \eq{3.5}. Comparing with \eq{2.6}, \eq{2.7} we find agreement provided that

\bea
 K_{ij}&=& -\frac{i}{2}\d_{ij} B :=-\frac{i}{2}\d_{ij} \L\L\nn\w1
 L_{a ij}&=&\, iA_{a ij}:= i\L \c_a \S_{ij} \L\nn\w1
 M_{ijkl}&=& -iB_{ijkl}:= -i\L\S_{ijkl}\L\ ,
 \la{3.8}
\eea

where, on the right-hand-side, the spacetime and internal spinor indices are contracted in the natural way (see appendix). The non-zero dimension-one torsion therefore becomes

\be
 T_{a\b j \c k}=-\frac{i}{2}(\c_a)_{\b\c}\d_{jk} B -i (\c_a{}^b)_{\b\c} A_{b jk}\ .
 \la{3.9}
\ee

With this, we now have a solution to the coupled Maurer-Cartan equations and geometrical Bianchi identities up to dimension one expressed entirely in terms of the physical fields. In terms of the sigma model fields the dimension-one curvatures are

\bea
 R_{\a i\b j, cd}&=& -\d_{ij}(\c_{cd})_{\a\b} B +2 \ve_{\a\b} A_{cdij}\ \Rightarrow\nn\w1
 R_{\a i,\b j,\c\d}&=&\half \d_{ij} (\c^a)_{\a\b} (\c_a)_{\c\d} B-
 \ve_{\a\b}(\c^a)_{\c\d} A_{a ij}\ ,\nn\w1
 R_{\a i\b j,kl}&=& \ve_{\a\b}(B_{ijkl} + 2 \d_{[i[k} \d_{j]l]}B)+(\c^a)_{\a\b} (4\d_{(i[k} A_{a j) l]}-\d_{ij} A_{a kl})\ .
 \la{3.9.1}
\eea

Note that there is an interesting feature of this solution that does not occur in higher-dimensional maximal supergravity theories (except for $D=11$), namely the fact that the dimension one-half torsion tensor is zero. This is easily understood in terms of group representations because in $D=3$ the spinor field transforms as a spinor under the internal symmetry group whereas the geometrical tensors can only accommodate tensor representations. If we move up to $N=8$ supergravity in $D=4$, for example, the internal symmetry group is $SU(8)$ and the spin one-half fermions transform under the $56$-dimensional representation. They can therefore be accommodated in the dimension one-half torsion as follows \cite{Brink:1979nt}:

\be
 T_{\a i,\b j, \cdt k}=\ve_{\a\b} \bar\L_{\cdt ijk} \ ,
 \la{3.10}
\ee

where we have used two-component spinor notation, where $i,j,k=1,\ldots 8$ and where $\bar\L_{\cdt ijk}$ is totally antisymmetric on its internal indices. Its leading component in a $\th$-exansion is the physical spin one-half fields in the 56 of $SU(8)$.


\section{Torsion and curvature up to dimension two}



\subsection{Dimension three-halves}


In this section we shall check the various identities up to dimension two. This will enable us to confirm the consistency of the solution and also to compute the dimension three-halves and two components of the torsion and curvature. As expected, these turn out to be functions of the physical fields, there being no gravitational degrees of freedom in three dimensions.

There are two relevant Bianchi identities, as well as the dimension three-halves components of $DP=0$ and $R=-P^2$. They are

\bea
 2 R_{[a \ua,b]c} &=& -T_{ab}{}^{\ub} T_{\ua\ub c} \la{4.1}\w1
 2 R_{a (\ua,\ub) \uc}&=& -2D_{(\ua} T_{a \ub) \uc} - T_{\ua\ub}{}^b T_{ab \uc}\la{4.2}\w1
 D_a P_{\ub}-D_{\ub} P_a + T_{a\ub}{}^{\uc} P_{\uc}&=& 0\la{4.3}\w1
 R_{a \b j, kl}&=& + 2iP_a  \S_{kl} \S_j \L_\a \ .
 \la{4.4}
\eea

Equation \eq{4.1} allows us to solve for the dimension three-halves Lorentz curvature in terms of the dimension three-halves torsion. The $\th=0$ component of the latter can be identified as the gravitino field strength, so we shall give it a new notation $T_{ab}{}^{\uc}:=\Psi_{ab}{}^{\uc}=\ve_{abc}\Psi^{c\,\uc}$. From \eq{4.2} we find that

\be
 \Psi_{a i}=2\c^b \c_a P_b \S_i \L\ ,
 \la{4.5}
\ee

confirming that the gravitino field strength is completely determined by the matter fields, as promised. The dimension-three-halves Lorentz curvature is

\be
R_{a\b j,b}=i(\c_a \Psi_{b i}-\half \h_{ab}\c^c  \Psi_{c i})_{\b}\ .
\la{4.5.1}
\ee

From \eq{4.3} we can determine the supersymmetry variation of $P_a$,

\be
 D_{\a i} P_a =i(\S_i D_a \L_\a)+ i T_{a\a i}{}^{\b j} (\S_j \L_\b)\ ,
 \la{4.6}
\ee

where the second term on the right-hand-side gives terms that are cubic in $\L$. 

At this stage we see that we have determined the geometric tensors in terms of the physical fields, but there are several other equalities that arise. One of these is the equation of motion for $\L$, and the others turn out to be identically satisfied even though this is not at all obvious at first sight. The full details of this are relegated to appendix D; here we simply state the equation of motion:

\be
 \c^a D_a \L=-\frac{i}{2} B\L +\frac{i}{6} \S^{ij} A_{a ij}\c^a\L. 
 \la{4.11}
\ee


\subsection{Dimension two}


There are two Bianchi identities at dimension two, the first of which simply tells us that the Riemann tensor $R_{ab,cd}$ has the usual symmetries in the absence of torsion. In three dimensions it can be written in the form

\be
 R_{ab,cd}=\ve_{abe}\ve_{cdf} G^{ef}
 \la{4.16}
\ee

where $G_{ab}:=R_{ab}-\half \h_{ab} R$ is the Einstein tensor. The second Bianchi identity is

\be
 R_{ab,\uc\ud}=2D_{[a} T_{b]\uc\ud} + D_{\uc} T_{ab\ud} + 2 T_{[a \uc}{}^{\ue} T_{b]\ue \ud}\ .
 \la{4.17}
\ee

In addition, we have the dimension-two component of the Maurer-Cartan equation which gives

\be
 R_{ab ij} = 2 P_a \S_{ij} P_b\ .
 \la{4.18}
\ee

It is a lengthy computation to analyse the content of these equations. Clearly, the $SO(16)$ curvature is immediately found from \eq{4.18}, while the Lorentz curvature is obtained from \eq{4.17}. But then there are a lot other components of \eq{4.17} which must be satisfied identically. It is indeed the case that this is so, but to prove it requires further Fierz rearrangement.

For the Lorentz curvature we find 

\be
 G_{ab}=-4(P_a P_b -\frac{1}{2}\h_{ab} P^c P_c) + 4i(\L \c_{(a} D_{b)}\L-\h_{ab} (B^2-\frac{1}{3} A_{a ij} A^{a ij}\ )) ,
 \la{4.19}
\ee 
 
where the right-hand side is essentially the on-shell energy-momentum tensor for the sigma model.

Finally, the equation of motion for the scalars can be found by acting on the fermion equation of motion with a spinorial derivative. It is

\be
 D^a P_a=\frac{i}{32} \S^{ij} P_a A^a_{ij}-\frac{i}{16. 6!}\S^{i_1\ldots i_6} P_a A^a_{i_1\ldots i_6}\ .
 \la{4.20}
\ee


\section{Forms}


In this section we discuss the various form field strengths that can arise in this theory. We shall concentrate on these rather than the potentials as this approach is gauge-invariant. This is particularly advantageous for the four-forms we shall discuss because the superspace field strengths can be non-vanishing even though the purely even components are identically zero. We shall derive the full set of forms up to degree four using only supersymmetry and $E_8$. In addition to the physical one-forms $P$ we are allowed to introduce their dual two-forms which transform under the 248 of $E_8$. Beyond these, we can in principle have three- and four-forms in arbitrary $E_8$ representations as the bosonic potentials do not introduce any new independent degrees of freedom; the problem is therefore to determine which representations are allowed. We also briefly mention the possible five-forms. The analysis of this problem is facilitated by the use of superspace cohomology techniques which we review in appendix C. The Bianchi identities for an $n$-form field strength have the following schematic form:

\be
I_{n+1}:=(d F_n - \sum F_p F_q)=0\ ; \quad p+q=n+1\ .
\la{5.1}
\ee

For the two-forms we have

\be
d F_2^R=0\ ,
\la{5.2}
\ee

where $R,S,T=1,\ldots 248$ denote adjoint representation indices. We shall see shortly that this identity is indeed satisfied, and that the $(2,0)$ component of $F_2$ is given by the dual of $P_a$ together with a bilinear fermion contribution. The components of the forms in an $E_8$ basis generically involve the scalar field matrix $\cV$ which we write, in the adjoint representation as,

\be
\cV_{\bar R}{}^R=\left(V_{ij}{}^R, V_I{}^R\right)\ ,
\la{5.3}
\ee

where the barred index is to be acted on by $SO(16)$ and therefore splits into the appropriate representations determined by the branching rules. The dimension-zero component of $F_2^R$ is given by

\be
F_{\a i\b j}^R=-2i\ve_{\a\b} V_{ij}{}^R\ .
\la{5.4}
\ee

Now suppose we have a set of three-forms $F_3^\cX$ transforming under some representation labelled by $\cX$. The Bianchi identity has the form

\be
d F_3^{\cX}=F_2^S F_2^R t_{RS}{}^{\cX}\ ,
\la{5.5}
\ee

where $t_{RS}{}^\cX$ is an $E_8$-invariant tensor. Because of \eq{5.2}, such a Bianchi is automatically consistent, in the sense that $d$ acting on both sides gives zero. The Bianchi identity is itself a four-form $I_4$ transforming in the representation $\cX$. Now any superspace $n$-form can be decomposed into a sum of $(p,q)$-forms, $p=n-q$, where $p\,(q)$ denotes the number of even (odd) indices carried by the form. The lowest-dimensional component of the four-form $I_4$ , $I_{0,4}$, has dimension zero, and since $d I_4=0$, it must satisfy $t_0 I_{0,4}=0$ ($t_0$ is defined in appendix C). Suppose that this equation is satisfied, then we will have $t_0 I_{1,3}=0$ at dimension one-half. But the cohomology group $H_t^{p,q}=0$ for $p\geq 1$ (see appendix C), so we have $I_{1,3}=t_0 J_{2,1}$. The equation $J_{2,1}=0$ can obviously be satisfied by an appropriate choice of $F_{2,1}$ because it contains exactly the right number of components, and so there can be no obstruction at dimension one-half. Similar arguments show that there are no higher-dimensional obstructions so that we conclude that the Bianchi identity \eq{5.5} is satisfied provided that its dimension-zero component is. 

The symmetric product of two $248$s is $1+ 3875 + 27000$. For the singlet the Bianchi identity is

\be
dF_3 =  F_2^R F_2^S a_{RS} \ ,
\la{5.6}
\ee

where $a_{RS}$ is the $E_8$ metric (given in appendix B).  The dimension-zero component is

\be
F_{a\b j\c k}=-i\d_{jk} (\c_a)_{\b\c}\ .
\la{5.7}
\ee

It is not difficult to see that $I_{0,4}=0$, and so we conclude that there is a singlet three-form.

Next consider the $3875$. The branching rule is $3875\rightarrow 135+1820 +1920'$. The 135 is a symmetric traceless tensor which we shall write as $t_{i,j}$, the $1820$ is a fourth-rank antisymmetric tensor, and the $1920'$ is a sigma-traceless primed vector-spinor. The dimension-zero component of this three-form is

\be
F_{a\b j\c k}^U=-i(\c_a)_{\b\c} V_{j,k}{}^U\ ,
\la{5.8}
\ee

where $U,V,W=1,\ldots 3875$ and where $\cV_{\bar U}{}^U$ is the scalar field matrix in the 3875 representation, so that $V_{i,j}{}^U$ is the projection onto the 135 in $\bar U$. If we write the Bianchi identity as $I_4^U=dF_3^U-F^S_2 F_2^R b_{RS}{}^U=0$, we can see that its dimension-zero component is indeed satisfied. This is because the symmetrised product of two $120$s coming from $F_{0,2}^S F_{0,2}^R b_{RS}{}^U$ can give both 135 and 1820 (since these are both contained in 3875), but the 1820 drops out in the Bianchi identity because the symmetrisation over the four odd indices would require antisymmetrisation over the four two-component Lorentz spinor indices. We can thus conclude without any further calculation that this Bianchi identity is satisfied.

The final possibility for three-forms is the 27000. The branching rule is $27000\rightarrow 1+1820+6435+5304+128+13312$, the last two being spinorial representations. The most significant one for us is the 5304; this is a tensor with the symmetries of the Weyl tensor (in sixteen dimensions). The only possibility for the dimension-zero component is 

\be
F^{\cX}_{a\b j\c k}=-i\d_{jk} (\c_a)_{\b\c} V_0{}^\cX\ ,
\la{5.9}
\ee

where $V_0{}^\cX$ denotes the singlet projection of the scalar matrix in the 27000. However, the dimension-zero $(F_2)^2$ term now has a contribution in the 5304 that cannot be balanced in the Bianchi identity, and so we conclude that the 27000 is not allowed.

Next, we consider the four-forms. They obey Bianchi identities of the form

\be
d F_4^\cX= F_3^U F_2^R t_{RU}{}^\cX\ ,
\la{5.10}
\ee

where $t_{RU}{}^\cX$ is an invariant tensor in the indicated representations. The possible representations are therefore contained in the tensor product of 248 and 3875 which is 779247+147250+30380+ 3875+248. In order for the Bianchi itself to be consistent we must have

\be
b_{(RS}{}^U t_{T)U}{}^\cX=0
\la{5.11}
\ee

This will be true if the symmetrised triple product of 248 does not contain the representation $\cX$. Of the possible representations, only the 3875 and the 147250 have this property and so we can discard the others. The Bianchi identities for the four-forms are five-forms, $I_5$, so that the lowest possible non-trivial components are $I_{1,4}$ (dimension zero). We must have $t_0 I_{1,4}=0$ which implies that $I_{1,4}=t_0 J_{2,2}$. Setting $J_{2,2}=0$ simply allows us to solve for the dimension-zero components of the four-forms, and the argument can be repeated at dimension one-half. There are no dimension-one components as we are in three-dimensional spacetime. We therefore conclude that there are no obstructions to solving any consistent (i.e. closed) Bianchi identities for four-forms.

There is a possible singlet four-form $F_4$ but it has to be gauge-trivial, i.e. $d F_4=0$. However, this is trivial in the sense that one can write $F_4=d G_3$, where the only non-zero component of $G$ is $G_{abc}\propto \ve_{abc}$.

Finally, we comment on the possible five-forms that can arise in the theory. The corresponding potentials for these do not have purely even components, but there are three-form gauge parameters that can have non-zero $(3,0)$-components. Indeed, as has been pointed out \cite{deWit:2008ta}, these can play a r\^{o}le in the gauged theory. The five-form Bianchi identities have the schematic form

\be
dF_5^{\cX}= (F_4 F_2)^\cX + (F_3 F_3)^\cX\  .
\la{5.11.1}
\ee

The possible representations are therefore contained in the products $248 \xz 3875=248+3875+30380+147250+779247$, $248\xz 147250=3875+30380+147250+779247+2450240+6696000+2641100$ and the antisymmetric product of two $3875$s which gives $248+30380+779247+6696000$. The only component of a five-form that can be non-zero in supergravity is $F_{3,2}$ and this must be proportional to $\ve_{abc}\ve_{\a\b}$ multiplied by a function of the form $F_{ij}^\cX$ that is antisymmetric on $ij$. It therefore follows that the only representations that can be non-zero must contain $120$ in the branching down to $SO(16)$. This leaves only four possibilities:$\,248, 30380,779247$ and $6696000$.  It is straightforward to verify, using group theory, that all four of these, which have contributions from at least two terms on the right-hand side, have consistent Bianchi identities and that they all occur with multiplicity one, i.e. there are no free parameters.  Given that the Bianchi identities are consistent, it follows from cohomology that there will be no obstructions to solving them.

To see this in more detail, consider an arbitrary five-form with the Bianchi identity \eq{5.11.1}. Applying another $d$ to this equation we find that both terms give rise to terms of the form

\be
F_3^U F_2^R F_2^S t_{U,RS,}{}^\cX\ ,
\la{5.11.2}
\ee

where $t$ is an invariant tensor, and the sum of these terms must vanish for consistency. The pair $RS$ is symmetric and can therefore be in any of the representations $1+3875+27000$. So to check whether a given term can be non-zero we have only to check if this representation multiplied by the $3875$ contains the representation $\cX$. Clearly the singlet cannot occur, but the other two can, at least in principle. As an example consider $\cX=R$, the 248. There are two terms in the Bianchi identity and the one coming from two three-forms can only give rise to $RS$ in the $3875$. So if the other term, coming from $F_4(3875)\xz F_2$ were to allow the 27000 the Bianchi identity would not be consistent. But $3875\xz 27000$ does not contain the 248, so the two $3875$ contributions can cancel and we are left with precisely one consistent Bianchi identity. The argument can easily be repeated for the other three representations that can tolerate non-zero five-forms in supergravity and one finds that they are indeed all consistent with multiplicity one.

The same argument can be extended to the representations corresponding to the five-form field strengths that must be zero in supergravity. They are $3875, 147250, 2450240$ and $26411008$. The  last two are definitely not consistent and  it is unlikely that other two are either, although this has not been checked in detail.

In summary, the dual two-forms are in the adjoint representation of $E_8$, the allowed three-forms transform under the singlet and 3875 representations and the allowed four-forms transform under the 3875 and 147250 representations together with a trivial singlet. The Bianchi identities are

\bea
dF_2^R&=&0 \la{5.12}\w1
dF_3&=& F_2^S F_2^R a_{RS}\la{5.13}\w1
dF_3^U&=&F_2^R F_2^S  b_{RS}{}^U\la{5.14} \w1
dF_4&=&0 \la{5.15}\w1
dF_4^U&=& F_3^V F_2^R c_{RV}{}^U\la{5.16}\w1
dF_4^X&=& F_3^V F_2^R d_{RV}{}^X\ ,
\la{5.17}
\eea

where $a,b,c,d$ are $E_8$-invariant tensors, and where $X,Y,Z=1,\dots 147250$.

The components of $F_2^R$ are 

\bea
F_{\a i\b j}^R&=& -2i\ve_{\a\b} V_{ij}{}^R\nn\w1
F_{a\b j}^R&=& -i(\c_a\S_j\L)_{\b }{}^I V_I{}^R\nn\w1
F_{ab}^R&=& \ve_{ab}{}^c (P_c{}^I V_I{}^R -2i A_c{}^{ij} V_{ij}{}^R)
\la{5.18}
\eea

The components of the singlet $F_3$ are

\bea
F_{a\b j\c k}&=&-i\d_{jk}(\c_a)_{\b\c}\nn\w1
F_{ab\c k}&=&0\nn\w1
F_{abc}&=& 4i\ve_{abc} B\ .
\la{5.19}
\eea

The components of $F_3^U$ are

\bea
F_{a\b j\c k}^U&=&-i(\c_a)_{\b\c} V_{j,k}{}^U\nn\w1
F_{ab\c k}^U&=& 2iV_{kI'} {}^U(\c_{ab}\L)_{\c I'}  \nn\w1
F_{abc}^U&=& 2i\ve_{abc} B^{ijkl} V_{ijkl}{}^U\ .
\la{5.20}
\eea

\black

The components of $F_4^U$ are

\bea
F_{ab \c k\d l}^U&=&-i(\c_{ab})_{\c\d} V_{k,l}{}^U\nn\w1
F_{abc \d l}^U&=& a\ve_{abc} V_{l I'} {}^U \L_{\c I'}  \ ,
\la{5.21}
\eea

while the components of $F_4^X$ are 

\bea
F_{ab \c k\d l}^X&=&-i(\c_{ab})_{\c\d} V_{k,l}{}^X\nn\w1
F_{abc \d l}^X&=& b \ve_{abc} V_{l I'} {}^X \L_{\c I'}  \  ,
\la{5.22}
\eea

where $a,b$ are real, calculable constants. 
It is easy to see that the singlet four-form $F_4$ is exact as the only non-zero component is

\be
F_{ab\c k\d l}=-i\d_{kl} (\c_{ab})_{\c\d}\ .
\la{5.23}
\ee

Clearly $F_4=d G_3$,  where the only non-vanishing component of $G_3$ is $G_{abc}=\ve_{abc}$. 

In addition there can be non-zero five-forms in the representations 248,30380,779247 and 6696000, obeying Bianchi identities of the form \eq{5.11.1}. The five-forms can only be non-vanishing at dimension zero where they have expressions of the form

\be
F_{abc\a i\b j}^\cX=i c\ve_{abc}\ve_{\a\b} V_{ij}{}^\cX\ ,
\la{5.24}
\ee

where $\cX$ can be one of the above representations, $ij$ denotes the 120 of $SO(16)$ and $c$ is some real constant.

The forms can equally well be discussed in an $SO(16)$ basis. We shall distinguish this basis by barring quantities or indices. The Bianchi identities can be written

\be
D \bar F_n =- \bar F_n\wedge P + \bar F_{n-1} \wedge \bar F_2
\la{5.25}
\ee

where $\bar F= F \cV$, $P$ is considered as being Lie-algebra-valued in the appropriate representation (with barred indices) and where the last term is understood as involving the appropriate invariant tensor.  For each $F$ this equation can be split into various representations of $SO(16)$ according to the branching rules. The components of the $F$s in this basis can be read off from those of the $E_8$ basis straightforwardly. A key point is that they do not contain any explicit scalars; in particular, the dimension-zero components are just given by $SO(16)$-invariant tensors.


\section{Gauging}



\subsection{Geometry}


The gauging of maximal $D=3$ supergravity has been discussed in \cite{Nicolai:2000sc,Nicolai:2001sv}, and the differential forms were subsequently discussed in \cite{deWit:2008ta}. The key tool is the embedding tensor, $\cE_R{}^S$.\footnote{The embedding tensor is usually called $\Th$ but we have chosen a different notation to avoid confusion with the superspace coordinates.}  The embedding tensor allows one to present the results in a way which looks $E_8$ covariant but which is actually only covariant with respect to the local $SO(16)$ and the gauge group $G_0\subset E_8$ that we shall not need to specify explicitly (see \cite{Nicolai:2001sv} for a list of the possible gauge groups). The embedding tensor is essentially given by a sum of projectors onto the irreducible subspaces of $\ge_8$ corresponding to the simple factors of $\gg_0$ \cite{Nicolai:2001sv}. It can be taken to be symmetric, $\cE_{RS}:=\cE_R{}^T a_{TS}=\cE_{SR}$ and there is also a quadratic constraint on $\cE$ that follows from demanding that it be invariant under gauge transformations. It is

\be
\cE_R{}^P \cE_{(M}{}^Q f_{N)PQ}=0\ ,
\la{6.0}
\ee

where $f_{PQR}$ denotes the $\ge_8$ structure constants. The discussion is best approached via the gauged Maurer-Cartan form \cite{de Wit:1982ig} (see \cite{Howe:1981tp} for the superspace version) which can be written

\be
\F=\cD\cV \cV^{-1}=P+ Q\ ,
\la{6.1}
\ee

where $\cD$ is a gauge-covariant derivative that acts on the $E_8$ index carried by $\cV_{\bar R}{}^R$, i.e. the superscript. The gauged Maurer-Cartan equation, which follows directly from \eq{6.1}, is

\be
R+ DP+ P^2 = g\cF:=g\cV \cF\cV^{-1}\ .
\la{6.2}
\ee

Here, $g$ is a constant with dimensions of mass which characterises the deformation and  $D$ is covariant with respect to both $SO(16)$ and $G_0$. The theory has both of these groups as local symmetries, but the rigid $E_8$ is broken. The technique we shall use in the following analysis is to work with $SO(16)$ indices, so that the gauge group is hidden from view.

The original geometrical constraint in superspace \eq{2.1}, i.e. taking the dimension-zero torsion to be the same as in flat space, together with the allowed conventional constraints, leads to the dimension-one torsion and curvatures given in equations \eq{2.6} and \eq{2.7}. Since the deformation parameter $g$ has dimension one it follows that we can expect changes to the tensors $K_{ij}, L_{a ij}$ and $M_{ijkl}$. These can only be proportional to $g$ multiplied by functions of the scalars and so $L_{a ij}$ must be unchanged. This leaves $K$ and $M$ which together fall into the $1+135+1820$ representations of $SO(16)$. Anticipating a little we can see that these can be combined into the $E_8$ representations $1+3875$ if we also have the $1920'$. This representation can be found in the scalar part of $D_{\a i}\L_{\b I'}$, which vanishes in the non-gauged case but which will be modified when the gauging is turned on as one can see from \eq{6.2}.

To implement the gauging explicitly we first need  to solve for the two-form field strength. This should be projected along $\gg_0$  which leads us to propose that it should have the form

\be
\cF^R=F^S\cE_S{}^R\ .
\la{6.2.1}
\ee

It is easy to see, using the fact that $\cD\cE_R{}^S=0$, that the Bianchi identity for $\cF^R$ will be solved if we take the components of $F^R$ to have the same form as in the ungauged case. In fact, the only $g$-dependence could be at dimension one, but since this component of $F$ is a spacetime two-form this cannot occur. Lowering the index on $\cF$ and converting to an $SO(16)$ basis we find, at dimension zero,

\bea
\cF_{\a i\b j,kl}&=&-2i\ve_{\a\b} V_{ij}{}^R V_{kl}^S\, \cE_{RS}\w1
\la{6.2.2}
\cF_{\a i\b j,I}&=&-2i\ve_{\a\b}  V_I{}^R V_{kl}^S\, \cE_{RS}\ , 
\la{6.3}
\eea

and at dimension one-half,
\be
\cF_{a\b j,I}=-i(\c_a \S_j\L)_\b{}^J  V_J{}^R V_{I}^S\, \cE_{RS}\ .
\ee

Since $\cE_{RS}$ is symmetric it can contain the $1+3875+27000$ representations of $E_8$, but we can see directly from the 120 component of \eq{6.2} that the $27000$ must be absent due to the fact that it cannot be accommodated in the dimension-one curvature. This is the basic constraint on $\cE$ derived  in \cite{Nicolai:2000sc,Nicolai:2001sv}. It then follows that the only representation in \eq{6.3} will be the 1920'. The functions appearing in the dimension-zero and one-half $\cF_2$'s can therefore be written

\bea
V_{ij}{}^R V_{kl}^S\, \cE_{RS}&=&f_{ij,kl}:=\d_{i[k}\d_{l]j} f_0 + \d_{[i[k} f_{j],l]} + f_{ijkl}\nn\w1
 V_I{}^R V_{kl}^S \cE_{RS}\,&=& (\S_{[k})_{IJ'} f_{l] J'}\nn\w1
V_{I}{}^R V_{J}^S\, \cE_{RS}&=&\frac{1}{4!}(\S_{ijkl})_{IJ}f^{ijkl}-2\d_{IJ}f_0\ , 
\la{6.4}
\eea

where the functions $f_0,f_{i,j},f_{ijkl}$ and $f_i$ exhibit the $1+135+1820+1920'$ split explicitly. The deformation feeds into the geometrical tensors at dimension one via the gauged Maurer-Cartan equation from which we find
 
\bea
K_{ij}&=& -\frac{i}{2}\d_{ij}B +2g( f_{i,j}+\d_{ij} f_0)\nn\w1
M_{ijkl}&=& -i B_{ijkl} +8g f_{ijkl}\nn\w1
D_{\a i}\L_{\b I'}&=&(\c^a)_{\a\b} (\S_i P_a)_{I'} +g\ve_{\a\b} f_{i I'}\ .
\la{6.5}
\eea

It is easy to check that the geometrical Bianchi identities at dimension three-halves are satisfied. To do this one needs the following easily derivable identities

\bea
D_{\a i}f_{0}&=&0\nn\w1
D_{\a i}f_{j,k}&=& 2i\L_{\a}\S_i\S_{(j}f_{k)}\nn\w1
D_{\a i}f_{jklm}&=& -i\L_{\a}\S_i\S_{[jkl}f_{m]}\nn\w1
D_{\a i}f_{j}&=& i\L_{\a}\S_i\S^k f_{j,k} + 
               \frac{i}{48}\L_{\a}\S_i( \S_{j}{}^{klmn} f_{klmn} +12 \S^{klm}f_{jklm})\ .
\la{6.5.0}
\eea 

There is a modification to the gravitino field strength given by
\be
\Psi_{a i}(g)=4g f_{i} \c_a \L\ ,
\la{6.5.1}
\ee

as well as a $g$-dependent term in the fermion equation of motion,

\be
\c^a D_a \L(g)=4g(f_0 \L -\frac{1}{48}f_{ijkl} \S^{ijkl}\L)\ .
\la{6.5.2}
\ee

Finally, at dimension two, there are changes to the curvature scalar and the scalar equation of motion given by

\bea
R(g)&=&-\frac{2ig}{3}(48f_0 B - f_{ijkl} B^{ijkl})\nn\\
    &&-3g^2(f_if^i-2f_{i,j}f^{i,j}-32f_0f_0)\nn\w1
D^a P_a(g)&=& \frac{5ig}{4.4!}\S_{ijk} f_l B^{ijkl}\nn \w1 
&&-\frac{3g^2}{4}(\S_if_jf^{i,j}-\frac{1}{9}\S_{jkl}f_i f^{ijkl})\, .
\la{6.5.3}
\eea


\subsection{Forms}


We now consider the hierarchy of forms. In the geometrical discussion above we have only used the true non-abelian gauge fields, but in order to accommodate all of the forms it will be necessary to include the other two-form gauge fields which we could think of as being abelian, although they do transform under the gauge group. In other words we have a set of 248 gauge fields $F^R$, where $\cF^R=F^S\cE_S{}^R$. The Bianchis for the forms can then be written, in the $E_8$ basis,

\be
\cD F_n =(F F)_{n+1}+ g F_{n+1}Y_{n+1,n}\ ,
\la{6.6}
\ee

where $(FF)$ denotes the bilinear term of the same form as in the ungauged case and $Y_{n+1,n}$ denotes a mapping from the representation space $\cR_{n+1}$ of the $(n+1)$-forms to that of the $n$-forms, $\cR_n$. In order to determine the $Y$-matrices one must compute the effect of applying $\cD$ to \eq{6.6}; clearly one will require, in agreement with the general discussion in \cite{deWit:2008gc}, that 

\be
Y_{n+1,n}\ Y_{n,n-1}=0
\la{6.7}
\ee

in order for the $g^2$ terms to cancel. The presence of the $g$-dependent term on the right-hand side of a deformed Bianchi identity implies that the $(n-1)$-form potential will transform under the $(n-1)$-form gauge transformation of the $n$-form potential, which is the way the hierarchy has been derived previously \cite{deWit:2008ta}. Before discussing this system in more detail we note that the $F$s themselves are hardly changed from the abelian case. Since $g$ has dimension one, it is only the purely even components of the $F$s that can get deformed and these only by the $f$-functions of the previous section. So only the $(3,0)$ components of the three-forms can receive corrections, which for the 3875 take the form

\be
F_{abc}^U=\ve_{abc} (a' f_0 V_0{}^U + b' f^{i,j} V_{i,j}{}^U + c' f^{ijkl} V_{ijkl}{}^U)\ ,
\la{6.8}
\ee

where $a',b'.c'$ are constants.  

We now give an example of the hierarchy computation in this covariant language. The two-form Bianchi identity is $\cD F_2=gF_3 Y_{3,2}$. Applying a second $\cD$ to this we get 

\be
gF_2^T F_2^S X_{ST}{}^R=gF_2^S F_2^T b_{ST}{}^U Y_U{}^R + g^2 F_4^{\cX} Y_{\cX}{}^U Y_U{}^R\ ,
\la{6.9}
\ee

where $X_{ST}{}^R=\cE_S{}^P f_{PT}{}^R$ is the generator of the gauge group within the 248 representation \cite{deWit:2008gc}.

The second term on the right must vanish in order to satisfy \eq{6.7}, and we can easily satisfy the part of the equation linear in $g$ by taking

\be
Y_{ST}{}^R=\cE_{(S{}}{}^P f_{T)P}{}^R
\la{6.10}
\ee

in agreement with \cite{deWit:2008ta}. Here, we have replaced $U$ by a symmetrised pair of $248$ indices on $Y$, and \eq{6.10} is correct as it stands because the singlet and the 27000 vanish on the right. One can continue in this way for the higher forms for which we find $Y_{4,3}$s that agree with those of \cite{deWit:2008ta}, although we have not checked beyond this. The complete hierarchy for the gauged forms in supergravity requires the five-forms and their Bianchi identities, and  the Bianchi identities for the six-forms, even though the latter vanish in supergravity. This is because the seven-form right-hand side of the six-form Bianchi identities involve terms of the form $(F_5,F_2)$ and $(F_4,F_3)$ and these expressions can in principle be non-zero at dimension zero. However, provided that the identities are themselves consistent, these equations will automatically be satisfied for cohomological reasons.

The fact that the Bianchi identities are consistent suggest that the field strengths should be expressible in terms of potentials, and this is indeed the case. For the two- and three-forms we find

\bea
F_2^R&=& d A_1^R + \frac{g}{2} A_1^T A_1^S X_{ST}{}^R + A_2^U Y_U{}^R\la{6.11}\w1
F_3^{RS}&=& \cD A_2^{RS} + g A_1^R( dA_1^S + \frac{1}{3} A_I^Q A_1^P X_{PQ}{}^S) + g A_4^\cX Y_\cX{}^{RS}\ ,
\la{6.12}
\eea

where, in the second equation, the indices $RS$ are symmetrised and projected onto the $3875$, while $\cX$ denotes the 3875 and the 147250 representations.

In the above we have ignored the singlet three-form, but its Bianchi identity can also be deformed:

\be 
dF_3=F_2^R F_2^S a_{RS} + g F_4^U Y_U\ ,
\la{6.13}
\ee

where $Y_U$ is proportional to the 3875 component of the embedding matrix. It is not difficult to verify the consistency of this Bianchi identity.


\section{Conclusions}


In this paper we have presented the maximal supergravity theory in three dimensions in a superspace setting. Starting from the off-shell superconformal constraints we have shown how one can accommodate the supergravity sigma model by introducing the scalar fields in the coset $SO(16)\bsh E_8$. The sigma model fields enter the geometry via the dimension-one components of the torsion and curvature tensors, that is the functions $K_{ij}, L_{a ij}$ and $M_{ijkl}$. The spin one-half fields $\L$ cannot appear in the dimension one-half torsion owing to the fact that they transform under a spinor representation rather than a tensor one, although this does not imply that there are Lorentz-covariant CR structures of a higher rank than there are in more spacetime dimensions (appendix E). We solved for all of the components of the torsion and curvature tensors and derived the equations of motion. 

The theory can be gauged by introducing a non-abelian gauged subgroup $G_0$ of $E_8$ and making use of the gauged Maurer-Cartan form. There are terms in the gauge-deformed Maurer-Cartan equation involving the two-form gauge field strength that are proportional to the parameter $g$ and that modify the dimension-one scalar functions in the theory. We computed these, the changes induced in all the components of the geometrical tensors and the modifications to the equations of motion. The dimension-one functions fall into the representations $1+135+1820+1920'$ of $SO(16)$, the representations that appear in the $1+3875$ representations of $E_8$. This gives a nice derivation of the fact that the embedding tensor, which is in the symmetric product of two adjoint representations of $E_8$, cannot contain the 27000 representation. This picture extends to higher dimensions. For example, in $D=4$ there are dimension-one scalars that appear in the torsion in the $36+420$ representations of $SU(8)$ \cite{Howe:1981gz} that can be combined into the $912$ representation of $E_7$, as one would expect \cite{deWit:1983gs}. 

The ungauged theory admits a set of differential forms that transform in various representations of $E_8$, including one-, two- and three-form potentials. We examined which representations can appear using supersymmetry and consistency of the Bianchi identities. The analysis is made easier in superspace owing to the fact that one can study the problem covariantly using the field strengths even in the case of the three-form potentials, because a four-form makes sense in superspace even when there are only three even dimensions. In addition, explicitly checking that the Bianchi identities are satisfied is facilitated by the use of superspace cohomology; only a few components of the Bianchi identities need to be checked due to the fact that $H_t^{p,q}=0$ for $p\neq 0$ in $D=3,N=16$ superspace.

The differential forms can also be studied in the gauged case by deforming the Bianchi identities. The field-strengths transform under the gauge group so that the exterior derivative must be made  gauge-covariant. The Bianchi identity for $F_n$ develops a term $gF_{n+1}Y_{n+1,n}$, where $Y_{n+1,n}$ maps the representation space of $(n+1)$-forms to that of $n$-forms and depends linearly on the embedding tensor. This means that all of the forms become related by a sequence of such maps that must be exact in order for the Bianchi identities to be consistent.   The closure of the full system of forms requires the introduction of five-forms in the supergravity limit, and it was shown that there are  such objects in the ungauged case. These are unmodified in the gauged case, and indeed the only forms that can be are the three-forms at dimension zero.

\pagebreak

{\bf\Large{Acknowledgements}}

We thank Eric Bergshoeff, Fabio Riccioni, Duncan Steele and Elias Lousseief for helpful comments.

JG thanks Tekn. Dr Marcus Wallenbergs Stiftelse and the STFC for financial support.

\vskip .5cm

 \appendix
 
 {\bf \Large{Appendices}}
 
 
\section{Spacetime coventions}


The metric is $\h_{ab}={\rm diag}(-1,1,1)$. The epsilon tensor is defined so that $\ve_{012}=+1$. The dual of a one-form $v_a$ is $v_{ab}:=\ve_{abc} v^c$ so that $v_a=-\frac{1}{2}\ve_{abc} v^{bc}$.

The gamma-matrices with indices in standard position are $(\c^a)_\a{}^\b$. They obey the algebra $\c_a \c_b =\h_{ab} + \c_{ab}$, where $\c_{ab}=\ve_{abc}\c^c$. We also have $\c_{abc}=\ve_{abc}$ for the totally antisymmetrised product of three gamma-matrices. Spinor indices are lowered or raised with the spin ``metrics'' $\ve_{\a\b}$ and $\ve^{\a\b}$ which we take to have the same numerical entries, i.e. $\ve_{12}=\ve^{12}=+1$. The summation convention is NE-SW, i.e. $v^\a=\ve^{\a\b} v_\b$ and $v_\a=v^\b\ve_{\b\a}$. The matrices $\c_a$ (and $\c_{ab}$) with both spinor indices down (or up) are symmetric.

A vector can be written as a symmetric bi-spinor via

\be
 v_{\a\b}=-\half (\c_a)_{\a\b} v_a \Leftrightarrow v_a= (\c_a)^{\a\b} v_{\a\b} \   .
 \la{a1}
\ee

For any two spinors $\psi,\chi$ and any gamma-matrix $\C$ we define the tensorial bilinear to be

\be
 \psi\C\chi:= \psi^\a \C_\a{}^\b \chi_\b\  .
 \la{a2}
\ee

 
\section{Conventions for $SO(16)$ and $E_8$}.


Vector indices are $i,j,\ldots =1\ldots 16$, unprimed Weyl spinor indices are $I,J,\ldots= 1\ldots 128$ and primed Weyl spinor indices are $I',J',\ldots =1\ldots 128$. The metrics for each three spaces are flat euclidean, so it is not important to distinguish between upper and lower indices.

The basic sigma-matrices are $(\S_i)_{IJ'}$ and $(\tilde\S_i)_{J'I}$. We shall take $\tilde\S_i=(\S_i)^T$ and not bother to write out the tildes since it will be clear from the context which is meant. Sigma-matrices with two or more indices are antisymmetrised products of the basic ones as usual.

Sigma-matrices with an even number of vector indices are bi-spinors of a fixed chirality. $(\S_2,\S_6)$ give a basis of antisymmetric $128\xz 128$ matrices while $(1,\S_4,\S_8)$ give a basis of symmetric matrices. We shall take $(\S_{i_1\dots i_8})_{IJ}$ to be self-dual while $\S_8$ with primed indices is anti-self-dual.

For an arbitrary matrix $M_{IJ}$ we have

\be
 M_{IJ} =\frac{1}{128} \sum_{n=0}^{n=4} (\S^{i_1\dots i_{2n}})_{IJ} M_{i_1\dots i_{2n}}\ ,
 \la{a3}
\ee

where 

\be
 M_{i_1\dots i_{2n}}:= \frac{1}{(2n)!}(\S_{i_1\dots i_{2n}})^{IJ} M_{IJ}\ ,
 \la{a4}
\ee

except for $n=4$ when there is an extra factor of $\half$ on the right-hand side. The matrix $\S_0$ is  $\d_{IJ}$. The formula for primed indices is identical.

The bilinears that can be formed from the spinor field $\L_{\a I'}$  are the Lorentz scalars

\bea
B&=& \L\L:= \L^{\a I'} \L_{\a I'} \nn\w1
B_{i_1\ldots i_4}&=& \L\S_{i_1\dots i_4}\L:=\L^{\a I'} (\S_{i_1\dots i_4})_{I'J'} \L_{\a J'}\nn\w1
B_{i_1\ldots i_8}&=& \L\S_{i_1\dots i_8}\L:=\L^{\a I'} (\S_{i_1\dots i_8})_{I'J'}\L_{\a J'} \ ,
\la{a5}
\eea

and the spacetime vectors

\bea
A_{a i_1 i_2}&=&\L \S_{i_1 i_2}\c_a \L:=\L^{\a I'} (\S_{i_1 i_2})_{I'J'} (\c_a)_\a{}^\b \L_{\b J'}  \nn\w1
A_{a i_1\ldots i_6}&=&\L \S_{i_1\dots i_6}\c_a \L:=\L^{\a I'} (\S_{i_1\dots i_6})_{I'J'} (\c_a)_\a{}^\b \L_{\b J'} \ .
\la{a6}
\eea

The adjoint representation of $E_8$ is the same as the defining representation and has dimension 248. It splits into 120+128 in $SO(16)$. The summation convention is $A^R B_R=A^I B_I +\half A^{ij} B_{ij}$. The metric $a_{RS}$ has components

\bea
a_{IJ}&=&\d_{IJ}\nn\w1
a_{ij,kl}&=&-\half\d_{k[i} \d_{j]l}\ .
\la{a7}
\eea

For the inverse, the summation convention implies that $a^{IJ}=\d^{IJ}$ while

\be
a^{ij,kl}=-8\d^{k[i} \d^{j]l}\ .
\la{a8}
\ee


\section{Superspace cohomology}


Since the tangent bundle splits into even and odd parts it is possible to split the space of $n$-forms into spaces of $(p,q)$-forms, $p+q=n$, where a $(p,q)$ form has $p$ even and $q$ odd indices:

\be
\O^{p,q}\ni \o_{p,q}=\frac{1}{p! q!} E^{\b_q}\ldots E^{\b_1} E^{a_p}\ldots E^{a_1} \o_{a_1\ldots a_p\b_1\ldots \b_q}\ ,
\la{c1}
\ee

where, in this appendix, spinor indices run from 1 to 32. The exterior derivative splits into four terms with different bidegrees:

\be
d=d_0+ d_1 + t_0 + t_1\ ,
\la{c2}
\ee

where the bidegrees are $(1,0),(0,1), (-1,2)$ and $(2,-1)$ respectively. The first two, $d_0$ and $d_1$, are essentially even and odd differential operators, while the other two are algebraic operators formed with the dimension-zero and dimension three-halves torsion respectively. In particular,

\be
(t_0 \o_{p,q})_{a_2\ldots a_p \b_1\ldots \b_q}\propto T_{(\b_1\b_2}{}^{a_1}\o_{a_1|a_2\ldots a_p|\b_3\ldots \b_{q+2})}\ .
\la{c3}
\ee

The equation $d^2=0$ splits into various parts according to their bidegrees amongst which one has

\bea
(t_0)^2&=& 0\la{c4}\w1
t_0 d_1 + d_1 t_0&=&0\la{c5}\w1
d_1^2 +t_0 d_0+ d_0 t_0&=&0\ .
\la{c6}
\eea

The first of these enables us the define the cohomology groups $H_t^{p,q}$, the space of $t_0$-closed $(p,q)$-forms modulo the exact ones \cite{Bonora:1986ix}. The other two then allow one to define the spinorial cohomology groups $H_s^{p,q}$, but we shall not need these in this paper. In ten and eleven dimensions these cohomology groups are related to spaces of pure spinors and pure spinor cohomology respectively  \cite{Howe:1991mf,Howe:1991bx,Berkovits:2002zk}.

In $D=3, N=16$ supergravity the dimension-zero torsion is given in equation \eq{2.1}. The associated $t_0$ turns out to have trivial cohomology for $p\geq 1$, a result that greatly simplifies the problem of finding solutions to the differential form Bianchi identities. It can be derived by dimensional reduction from $D=10$ \cite{Berkovits:2008qw} cohomology.


\section{A dimension three-halves calculation}


In this appendix we give some more details of the dimension three-halves equations that arise in section 4. We can study these by computing two spinorial covariant derivatives acting on $\L$. We have

\be
 D_{\ua} D_{\ub}= D^2_{\ua\ub} + \frac{1}{2} \{D_{\ua}, D_{\ub}\}\ .
 \la{4.7}
\ee

The second-order derivative can be decomposed as

\be
 D^2_{\ua\ub}:= \ve_{\a\b} D^2_{ij} + (\c^a)_{\a\b} D^2_{a ij}
 \la{4.8}
\ee

where the Lorentz scalar and vector terms are respectively symmetric and antisymmetric on $ij$.  We can now evaluate two spinorial derivatives on $\L$ using this and the Ricci identity that enables us to express the anti-commutator in terms of the torsion times a single derivative and the curvature. We find

\bea
 \ve_{\a\b} D^2_{ij}\L_\c  + (\c^a)_{\a\b} D^2_{a ij} \L_\c 
 &=&-\frac{i}{2}\d_{ij} (\c^a)_{\a\b} D_a \L_\c+\frac{i}{2} (\c^a)_{\b\c} (\S_j \S_i D_a \L_\a)\nn\w1
 &\phantom{=}& + \frac{i}{2} (\c^a)_{\b\c}T_{a\a i}{}^{\d l} (\S_j\S_l\L_\d)\nn\w1
 &\phantom{=}& -\half \ve_{(\a|\c|}\L_{\b)} B -\half \ve_{\a\b} (\c^a\L)_{\c} A_{a ij}\nn\w1
 &\phantom{=}& +\frac{1}{8}(B_{ijkl} + 2 \d_{i[k} \d_{l]j}B) (\S^{kl} \L_\c)\nn\w1
 &\phantom{=}& + \frac{1}{8}(\c^a)_{\a\b} (4 \d_{(i[k} A_{a j) l]} -\d_{ij} A_{a kl}) (\S^{kl} \L_\c)\ .
 \la{4.9}
\eea

We now split this equation into four parts according to the symmetries of the pairs of spinor and internal indices. Consider first the part that is symmetric on $\a\b$ and on $ij$. After contracting the expression with $(\c_a)^{\a\b}$ and with a little algebra one can show that this is proportional to $\d_{ij}$. One finds

\be
 2i \c_a \c^b D_b \L - \c_a B\L +\S^{ij} A_{a ij} \L=0\ ,
 \la{4.10}
\ee

Contracting this with $\c^a$ we obtain the Dirac equation for $\L$,

\be
 \c^a D_a \L=-\frac{i}{2} B\L +\frac{i}{6} \S^{ij} A_{a ij}\c^a\L. 
 \la{4.11}
\ee

The gamma-traceless part of \eq{4.10} must therefore vanish identically. That it does so is due to the identity

\be
 \S^{ij} \L_{(\a} \L_{\b} \S_{ij} \L_{\c)}=0\ .
 \la{4.12}
\ee

The part that is symmetric on $\a\b$ and antisymmetric on $ij$ determines $D^2_{aij}\L$ to be

\be
 D^2_{a ij} \L=-\frac{i}{4} \c^b \c_a D_b\L -\frac{i}{4} \S_{[i}\S^k \c^b\c_a T_{b j]k}\L\ ,
 \la{4.13}
\ee

where we have regarded the dimension-one torsion as a matrix in spin space. In likewise fashion, the antisymmetric-symmetric part determines $D^2_{ij}\L$ to be

\be
 D^2_{ij}\L=\d_{ij}(\frac{7}{8} B\L +\frac{1}{12} \S^{kl} A_{kl}\L) -2 \S_{(i}\S^k A_{j)k}\L\ ,
 \la{4.14}
\ee    
  
where $A_{ij}:=-\half \c^a A_{a ij}$ is regarded as a matrix in spin space. Finally, we are left with the part that is antisymmetric on both pairs of indices. After making use of the equation of motion we obtain

\be
\S_{ij} B\L-\frac{1}{2} \S^{kl} B_{ijkl}\L-4A_{ij}\L -4\S_{[i} \S^k A_{j]k} \L +\frac{1}{3} \S_{ij} \S^{kl} A_{kl} \L=0\ .
 \la{4.15}
\ee

This equation, cubic in $\L$, has the form of an antisymmetric tensor-spinor with respect to the $SO(16)$ indices. The fact that it is identically true can be shown using the Grassmann-odd nature of $\L$ together with some Fierz rearrangement.


\section{Harmonic superspace}


Harmonic superspaces were introduced in   \cite{Rosly:1982,Galperin:1984av,Karlhede:1984vr} and first studied in three dimensions in \cite{Zupnik:1988wa}. A complete classification of harmonic superspaces for three-dimensional supersymmetry was given in \cite{Howe:1994ms}. In this section we discuss harmonic superspaces for maximal supergravity in $D=3$. 

The basic idea of harmonic superspace is to introduce additional bosonic variables that parametrise a coset space of the R-symmetry group, in our case $SO(16)$. Following \cite{Galperin:1984av}  we shall use an equivariant formalism, i.e. we work on the group $G=SO(16)$ with fields whose dependence on the isotropy subgroup $H$ is fixed; this has the advantage that one can work globally without having to introduce local coordinates on the coset. We denote a group element by $u_{i'}{}^i$, where $G$ acts to the right on $i$ and $H$ acts to the left on the primed index. We can use $u$ to convert $G$-indices on superfields to $H$ ones. To differentiate with respect to the harmonic variables we use the right-invariant vector fields $D_{i'j'}$ which are antisymmetric and which obey the commutation relations of $\gs\go(16)$. We have

\be
D_{i'j'} u_{k'}{}^k= \d_{k'[i'} u_{j']}{}^k\ .
\la{4a.1}
\ee

We are interested in constructing CR structures on harmonic superspace. A CR structure is an involutive complex distribution,  $K$ say, such that $K\cap \bar K=0$. In other words we have to look for sets of independent complex vector fields that close under graded commutation. In flat harmonic superspace the appropriate cosets are the flag manifolds $\bbF_k:= (U(k) \xz SO(16-2k))\bsh SO(16)$; they have the propoerties that they are both compact and complex. The fundamental representation space of $SO(16)$ is $\bbR^{16}$ which we can think of as $\bbR^{2k}\oplus \bbR^{16-2k}$. The isotropy group will then preserve this splitting together with a complex structure on $\bbR^{2k}$.  We set $i'=(r,\bar r, x)$, where $r=1\ldots k$ and $x=2k+1\ldots 16$. Note that in this partially complex basis the $SO(16)$ metric is $(\d_{r\bar s}, \d_{xy})$ so that the flat superspace derivatives $D_{\a r}=u_r{}^i D_{\a i}$ anti-commute. The derivatives on $G$ split into the isotropy group ones $(D_{r\bar s}, D_{xy})$, and two sets of coset derivatives that correspond to the $\bar\del$-operator and its complex conjugate. We have $\bar\del \sim (D_{rs}, D_{r y})$ and $\del \sim (D_{\bar r\bar s}, D_{\bar r y})$. Note that the set of derivatives $(D_{\a r}, D_{rs}, D_{r y})$ closes on itself and thus defines a CR structure. It is this sort of structure that we want to generalise to the curved superspace case.

In maximal $D=3$ supergravity,  harmonic superspace is the $H\bsh G$ bundle associated with the principal $SO(16)$ bundle. In order to construct the required CR structures we start by considering the horizontal lifts of the odd basis vector fields $E_{\a i}$,

\be
\wt E_{\a i'}:=E_{\a i'}-\half \O_{\a i'}{}^{j'k'} D_{j'k'}\ ,
\la{4a.2}
\ee

where $E_{\a i'}:= u_{i'}{}^i E_{\a i}$, etc. The anti-commutator of two such vector fields is

\be
[\wt E_{\a i'}, \wt E_{\b j'}]=\O_{\a i',\b}{}^{\c} \wt E_{\c j'} + \O_{\b j',\a}{}^{\c} \wt E_{\c i'}-T_{\a i',\b j'}{}^C \wt E_C + \half R_{\a i',\b j'}{}^{k'l'}D_{k'l'}\ ,
\la{4a.3}
\ee

where $\wt E_c$ is defined in a simliar way to $\wt E_{\a i'}$. 

We now wish to examine the CR structures spanned by sets of vector fields of the form $(\wt E_{\a r}, D_{rs}, D_{r x})$ that are compatible with the supergravity constraints. At first sight it might seem that the $D=3$ theory could be better behaved in this regard than maximal supergravity theories in higher dimensions because the dimension one-half torsion vanishes. We recall that in $D=4, N=8$ supergravity, the largest CR structure that is allowed has one two-component dotted and one two-component undotted odd vector field,  $(\wt E_{\a 1}, \wt {{E}}_{\adt}^8)$ say, which means that we can only have Lorentz covariant sub-superspace measures corresponding to integrals over at least twenty-eight odd coordinates \cite{Hartwell:1994rp,Bossard:2010bd}. However, we shall see that there is no improvement on this  in three dimensions due to the presence of the curvature term in \eq{4a.3}.

In examining the sets of vector fields that span the CR structure we need not consider the derivatives along the isotropy sub-algebra directions, and we are allowed to have non-zero curvatures that contract with the harmonic derivatives corresponding to $\bar\del$ in the coset space. So the curvatures that are constrained have Lie algebra components $(rs, ry)$. We thus require

\bea
R_{\a r,\b s, tu}&=&0\nn\w1
R_{\a r,\b s, t y}&=&0\ .
\la{4a.4}
\eea

Looking at the third line of \eq{3.9.1} we see that the terms involving the singlet scalar $B$ and the vector $A_{a ij}$ drop out of these expressions due to the presence of the $SO(16)$ metric, but that the four-index scalar $B_{ijkl}$ does not. In order for both of equations \eq{4a.4} to be satisfied it is clear that $r$ can take at most two values, and so we conclude that  the largest odd dimension for a Lorentz-covariant CR structure is four, which again corresponds to the possibility of integrals over twenty-eight odd coordinates. The isotropy group for this structure is $U(2)\xz SO(12)$.  Finally, we note that we can have analytic fields of this type (i.e. annihilated by all of the CR vector fields) provided that they only carry charges under the $U(1)$ subgroup of the $U(2)$ factor of the isotropy group. Analytic fields that transform under the rest of this group are not allowed due to the curvature terms in the isotropy group directions in \eq{4a.3}.



\begin{thebibliography}{99}

\bibitem{Cremmer:1979up}
  E.~Cremmer and B.~Julia,
  ``The SO(8) Supergravity,''
  Nucl.\ Phys.\  B {\bf 159} (1979) 141.

\bibitem{Nicolai:1988jb}
  H.~Nicolai and N.~P.~Warner,
  ``The structure of N=16 supergravity in two dimensions,''
  Commun.\ Math.\ Phys.\  {\bf 125} (1989) 369.

\bibitem{Kleinschmidt:2004dy}
  A.~Kleinschmidt and H.~Nicolai,
  ``E(10) and SO(9,9) invariant supergravity,''
  JHEP {\bf 0407} (2004) 041
  [arXiv:hep-th/0407101].
\bibitem{West:2001as}
  P.~C.~West,
  ``E(11) and M theory,''
  Class.\ Quant.\ Grav.\  {\bf 18} (2001) 4443
  [arXiv:hep-th/0104081].
  
\bibitem{Marcus:1983hb}
  N.~Marcus and J.~H.~Schwarz,
  ``Three-Dimensional Supergravity Theories,''
  Nucl.\ Phys.\  B {\bf 228} (1983) 145.
  
  
  
\bibitem{Cremmer:1997ct}
  E.~Cremmer, B.~Julia, H.~Lu and C.~N.~Pope,
  ``Dualisation of dualities. I,''
  Nucl.\ Phys.\  B {\bf 523} (1998) 73
  [arXiv:hep-th/9710119].
  
\bibitem{Cremmer:1998px}
  E.~Cremmer, B.~Julia, H.~Lu and C.~N.~Pope,
  ``Dualisation of dualities. II: Twisted self-duality of doubled fields  and
  superdualities,''
  Nucl.\ Phys.\  B {\bf 535} (1998) 242
  [arXiv:hep-th/9806106].
  
\bibitem{HenryLabordere:2002dk}
  P.~Henry-Labordere, B.~Julia and L.~Paulot,
  ``Borcherds symmetries in M-theory,''
  JHEP {\bf 0204} (2002) 049
  [arXiv:hep-th/0203070].
  
\bibitem{Julia:1997cy}
  B.~L.~Julia,
  ``Dualities in the classical supergravity limits: Dualisations,  dualities
  and a detour via 4k+2 dimensions,''
  arXiv:hep-th/9805083.
  
\bibitem{Riccioni:2007au}
  F.~Riccioni and P.~C.~West,
  ``The E(11) origin of all maximal supergravities,''
  JHEP {\bf 0707} (2007) 063
  [arXiv:0705.0752 [hep-th]].
  
\bibitem{Bergshoeff:2007qi}
  E.~A.~Bergshoeff, I.~De Baetselier and T.~A.~Nutma,
  ``E(11) and the embedding tensor,''
  JHEP {\bf 0709} (2007) 047
  [arXiv:0705.1304 [hep-th]].
  
\bibitem{Riccioni:2009xr}
  F.~Riccioni, D.~Steele and P.~West,
  ``The E(11) origin of all maximal supergravities - the hierarchy of
  field-strengths,''
  JHEP {\bf 0909} (2009) 095
  [arXiv:0906.1177 [hep-th]].
  
\bibitem{Henneaux:2010ys}
  M.~Henneaux, B.~L.~Julia and J.~Levie,
  ``$E_{11}$, Borcherds algebras and maximal supergravity,''
  arXiv:1007.5241 [hep-th].
      
  
\bibitem{deWit:2008gc}
  B.~de Wit and H.~Samtleben,
  ``The end of the p-form hierarchy,''
  JHEP {\bf 0808} (2008) 015
  [arXiv:0805.4767 [hep-th]].
  

  

  
\bibitem{Bergshoeff:2007ma}
  E.~Bergshoeff, P.~S.~Howe, S.~Kerstan and L.~Wulff,
  ``Kappa-symmetric SL(2,R) covariant D-brane actions,''
  JHEP {\bf 0710} (2007) 050
  [arXiv:0708.2722 [hep-th]].
  
\bibitem{Bergshoeff:2010mv}
  E.~A.~Bergshoeff, J.~Hartong, P.~S.~Howe, T.~Ortin and F.~Riccioni,
  ``IIA/IIB Supergravity and Ten-forms,''
  JHEP {\bf 1005} (2010) 061
  [arXiv:1004.1348 [hep-th]].
  
\bibitem{Nicolai:2000sc}
  H.~Nicolai and H.~Samtleben,
  ``Maximal gauged supergravity in three dimensions,''
  Phys.\ Rev.\ Lett.\  {\bf 86} (2001) 1686
  [arXiv:hep-th/0010076].
  
     
  
\bibitem{Nicolai:2001sv}
  H.~Nicolai and H.~Samtleben,
  ``Compact and noncompact gauged maximal supergravities in three
  dimensions,''
  JHEP {\bf 0104} (2001) 022
  [arXiv:hep-th/0103032].
 
\bibitem{deWit:2008ta}
  B.~de Wit, H.~Nicolai and H.~Samtleben,
  ``Gauged Supergravities, Tensor Hierarchies, and M-Theory,''
  JHEP {\bf 0802} (2008) 044
  [arXiv:0801.1294 [hep-th]].
  
\bibitem{Bergshoeff:2008qd}
  E.~A.~Bergshoeff, O.~Hohm and T.~A.~Nutma,
  ``A Note on E(11) and Three-dimensional Gauged Supergravity,''
  JHEP {\bf 0805} (2008) 081
  [arXiv:0803.2989].
  
  
\bibitem{Howe:1995zm}
  P.~S.~Howe, J.~M.~Izquierdo, G.~Papadopoulos and P.~K.~Townsend,
  ``New supergravities with central charges and Killing spinors in
  (2+1)-dimensions,''
  Nucl.\ Phys.\  B {\bf 467} (1996) 183
  [arXiv:hep-th/9505032].
  
\bibitem{Howe:2004ib}
  P.~S.~Howe and E.~Sezgin,
  ``The Supermembrane revisited,''
  Class.\ Quant.\ Grav.\  {\bf 22} (2005) 2167
  [arXiv:hep-th/0412245].
  
  
\bibitem{Kuzenko:2011xg}
  S.~M.~Kuzenko, U.~Lindstrom and G.~Tartaglino-Mazzucchelli,
  ``Off-shell supergravity-matter couplings in three dimensions,''
  JHEP {\bf 1103} (2011) 120
  [arXiv:1101.4013 [hep-th]].
   
  
\bibitem{Cederwall:2011pu}
  M.~Cederwall, U.~Gran and B.~E.~W.~Nilsson,
  ``D=3, N=8 conformal supergravity and the Dragon window,''
  arXiv:1103.4530 [hep-th].
  
\bibitem{Dragon:1978nf}
  N.~Dragon,
  ``Torsion And Curvature In Extended Supergravity,''
  Z.\ Phys.\  C {\bf 2} (1979) 29.
     
  
\bibitem{Brink:1979nt}
  L.~Brink and P.~S.~Howe,
  ``The N=8 Supergravity In Superspace,''
  Phys.\ Lett.\  B {\bf 88} (1979) 268.
  
\bibitem{de Wit:1982ig}
  B.~de Wit and H.~Nicolai,
  ``N=8 Supergravity,''
  Nucl.\ Phys.\  B {\bf 208} (1982) 323.

  
\bibitem{Howe:1981tp}
  P.~S.~Howe and H.~Nicolai,
  ``Gauging N=8 Supergravity In Superspace,''
  Phys.\ Lett.\  B {\bf 109} (1982) 269.
  
\bibitem{Howe:1981gz}
  P.~S.~Howe,
  ``Supergravity In Superspace,''
  Nucl.\ Phys.\  B {\bf 199} (1982) 309.
  
  
\bibitem{deWit:1983gs}
  B.~de Wit and H.~Nicolai,
  ``The Parallelizing S(7) Torsion In Gauged N=8 Supergravity,''
  Nucl.\ Phys.\  B {\bf 231} (1984) 506.
  
\bibitem{Bonora:1986ix}
  L.~Bonora, P.~Pasti and M.~Tonin,
  ``Superspace Formulation Of 10-D Sugra+Sym Theory A La Green-Schwarz,''
  Phys.\ Lett.\  B {\bf 188} (1987) 335.
  
 
  
\bibitem{Howe:1991mf}
  P.~S.~Howe,
  ``Pure Spinors Lines In Superspace And Ten-Dimensional Supersymmetric
  Theories,''
  Phys.\ Lett.\  B {\bf 258} (1991) 141
  [Addendum-ibid.\  B {\bf 259} (1991) 511].

\bibitem{Howe:1991bx}
  P.~S.~Howe,
  ``Pure Spinors, Function Superspaces And Supergravity Theories In
  Ten-Dimensions And Eleven-Dimensions,''
  Phys.\ Lett.\  B {\bf 273} (1991) 90.
  
\bibitem{Berkovits:2002zk}
  N.~Berkovits,
  ``ICTP lectures on covariant quantization of the superstring,''
  arXiv:hep-th/0209059.
  
\bibitem{Berkovits:2008qw}
  N.~Berkovits and P.~S.~Howe,
  ``The cohomology of superspace, pure spinors and invariant integrals,''
  arXiv:0803.3024 [hep-th].
      
\bibitem{Rosly:1982}
 A.~A.~Rosly,
 ``Super Yang--Mills  constraints
 as integrability conditions,'' in {\it Proceedings of the International
 Seminar on Group Theoretical
 Methods in Physics},'' (Zvenigorod, USSR, 1982),
 M. A. Markov  (Ed.),
 Nauka, Moscow, 1983, Vol. 1, p. 263 (in Russian);
 English translation: in {\it Group Theoretical
 Methods in Physics},'' M. A. Markov, V. I. Man'ko
 and A. E. Shabad  (Eds.), Harwood Academic Publishers,
 London, Vol. 3, 1987, p. 587.

\bibitem{Galperin:1984av}
  A.~Galperin, E.~Ivanov, S.~Kalitsyn, V.~Ogievetsky and E.~Sokatchev,
  ``Unconstrained $\cN=2$ matter, Yang--Mills and supergravity theories in harmonic
  superspace,''
  Class.\ Quant.\ Grav.\  {\bf 1}, 469 (1984).
  
\bibitem{Karlhede:1984vr}
  A.~Karlhede, U.~Lindstrom and M.~Rocek,
  ``Selfinteracting tensor multiplets In $\cN=2$ superspace,''
  Phys.\ Lett.\  B {\bf 147},  297 (1984).
  
\bibitem{Zupnik:1988wa}
  B.~M.~Zupnik and D.~V.~Khetselius,
  ``Three-dimensional extended supersymmetry in harmonic superspace,''
  Sov.\ J.\ Nucl.\ Phys.\  {\bf 47} (1988) 730
  [Yad.\ Fiz.\  {\bf 47} (1988) 1147].
  
\bibitem{Howe:1994ms}
  P.~S.~Howe and M.~I.~Leeming,
  ``Harmonic superspaces in low dimensions,''
  Class.\ Quant.\ Grav.\  {\bf 11} (1994) 2843
  [arXiv:hep-th/9408062].
  
\bibitem{Hartwell:1994rp}
  G.~G.~Hartwell and P.~S.~Howe,
  ``(N, P, Q) Harmonic Superspace,''
  Int.\ J.\ Mod.\ Phys.\  A {\bf 10} (1995) 3901
  [arXiv:hep-th/9412147].
  
\bibitem{Bossard:2010bd}
  G.~Bossard, P.~S.~Howe and K.~S.~Stelle,
  ``On duality symmetries of supergravity invariants,''
  arXiv:1009.0743 [hep-th].
  
    
 \end{thebibliography}
\end{document}